\documentclass[a4paper,12pt]{article}
\pdfoutput=1
\usepackage{graphicx}
\usepackage{xcolor}
\usepackage{setspace}
\usepackage[title]{appendix}
\usepackage{inputenc}
\usepackage{ulem}
\usepackage{amsmath}
\usepackage{eurosym}
\usepackage{url}
\usepackage{authblk} 
\usepackage{epstopdf}
\usepackage{placeins}
\usepackage{tabularx}
\usepackage{longtable}
\usepackage{adjustbox}
\usepackage{booktabs}
\usepackage{bm}
\usepackage{float}
\usepackage[round]{natbib}
\usepackage{amsmath} 
\usepackage{amssymb} 
\usepackage{latexsym}
\usepackage{wasysym}
\usepackage{subfigure}
\usepackage[flushleft]{threeparttable}
\usepackage{color}
\usepackage{caption}

\begin{document}
	\title{Shifting Correlations: How Trade Policy Uncertainty Alters stock-T bill Relationships}
\author{Demetrio Lacava\footnote{Email: dlacava@unime.it}}
\date{}

\affil[]{Department of Economics, University of Messina, Messina, via dei Verdi, 75, 98122, Italy.}
\maketitle  

	\begin{abstract} 
		\noindent This paper examines how trade policy uncertainty influences the correlation between U.S. stock indices and short-term government bonds. The objective is to assess whether policy-related shocks, especially those linked to trade tensions, alter the traditional stock-T bill relationship and its implications for investors. We extend the Dynamic Conditional Correlation (DCC) framework by incorporating exogenous variables to account for external shocks. Three specifications are analyzed: one using the Trade Policy Uncertainty (TPU) index, one including a dummy variable reflecting presidential-cycle effects, and one combining both through an interaction term. The analysis is based on daily data for major U.S. stock indices and the 3-month Treasury bill. Results indicate that trade policy uncertainty exerts a significant effect on stock-T bill correlations. Moreover, its influence becomes stronger under specific political conditions, suggesting that political agendas can amplify the impact of trade-related shocks on financial markets. Crucially, augmenting the DCC framework with trade-policy-related variables improves also the economic relevance of correlation forecasts. Therefore, this study contributes to the literature by explicitly integrating policy-related uncertainty into correlation modeling through an augmented DCC framework. The findings provide new insights for portfolio allocation and risk management in environments characterized by heightened trade tensions.
		\end{abstract}
	\textbf{Paper type} - Research article.
	
	\section{Introduction}

It is widely recognized that correlations between stock and bond returns evolve over time in both sign and magnitude, playing a key role in asset allocation, diversification, and risk pricing. Recent research further emphasizes that the covariance between equities and bonds represents an important source of intertemporal risk affecting expected returns and portfolio hedging demand \citep{Perras:Wagner:2020}. Moreover, the dynamics of stock-T bill correlations are found to be sensitive to different market phases \citep{Selmi:Gupta:Kollias:Papadamou:2021}.  The development of the Dynamic Conditional Correlation model \citep[DCC,][]{Engle:2002b} has significantly advanced the modeling and forecasting of time-varying correlations, becoming a standard reference in financial econometrics. The DCC represents a successful refinement of earlier contributions, including the Constant Conditional Correlation model \citep[CCC,][]{Bollerslev:1990}, in which the variances follow univariate GARCH processes while correlations are assumed constant. This framework has the advantage of being estimable through a two-step procedure, which significantly reduces the number of parameters by estimating $N$ univariate GARCH models (where $N$ is the number of assets), but at the cost of imposing a total of $N(N-1)/2$ constant correlations, typically set equal to their sample counterparts. Such a restriction may be overly strong in environments characterized by time-varying market conditions.
By contrast, within the DCC framework correlations are allowed to evolve dynamically as functions of their own past values and standardized innovations, often interpreted as market “news”. Importantly, this dynamic structure preserves the computational tractability of the two-step approach while ensuring the positive definiteness of the conditional correlation matrix and mitigating the curse of dimensionality, which refers to the exponential growth in the number of parameters as the number of assets increases. This issue is particularly severe in fully parameterized multivariate GARCH models such as the VECH GARCH \citep{Bollerslev:Engle:Wooldridge:1988} and the BEKK \citep{Engle:Kroner:1995}.

In addition to lagged correlations and past market innovations, a growing literature suggests that correlation dynamics may also respond to observable exogenous factors that shape investors’ expectations and portfolio allocation decisions. In this respect, uncertainty related to economic policy represents a natural candidate for capturing external shocks that are not fully summarized by return innovations alone. Incorporating such information into correlation models may enhance forecast accuracy and improve the assessment of diversification and risk-management strategies. Previous research indicates that measures of economic policy uncertainty affect correlations between both oil and stock markets \citep{Fang:Chen:Yu:Xiong:2018} and between bonds and equities \citep{Fang:Yu:Li:2017}. More generally, stock-bond correlations tend to rise with inflation expectations and decline under heightened market uncertainty \citep{Andersson:Krylova:Vahamaa:2008}, and are also influenced by market liquidity conditions \citep{Baele:Bekaert:Inghelbrecht:2010}. Related evidence highlights that variations in equity–bond covariance are closely linked to portfolio reallocations between risky and safe assets, often associated with flight-to-quality episodes in periods of elevated uncertainty \citep{Perras:Wagner:2020}. Despite these contributions, relatively limited attention has been devoted to modeling how specific sources of policy-related uncertainty enter directly into the dynamics of conditional correlations.

While trade policy uncertainty is related to broader measures of policy uncertainty -- such as the Economic Policy Uncertainty \citep[EPU,][]{Baker:Bloom:Davis:2016} index -- it captures a more specific dimension associated with shocks to trade policy discussions, negotiations, and tariff-related tensions. In this sense, trade uncertainty would reflect a distinct source of uncertainty that may affect investors' expectations regarding international trade conditions and global economic activity. Our objective is not to isolate a causal channel but to assess whether observable trade-related uncertainty contains incremental information for correlation dynamics beyond return innovations.

This paper contributes to the literature by extending the Dynamic Conditional Correlation framework to explicitly incorporate observable policy-related uncertainty into the correlation dynamics. Rather than assuming that external shocks are fully embedded in return innovations, we allow trade policy uncertainty to directly affect the evolution of stock-T bill correlations through an augmented DCC specification. This approach preserves the parsimony and computational tractability of the standard DCC model, while enhancing its economic interpretability. To the best of our knowledge, this paper is among the first to integrate trade policy uncertainty into a DCC-type model, providing a transparent framework to assess how trade-policy shocks shape cross-asset comovements. Accordingly, the main contribution of the paper lies in highlighting the role of trade-related uncertainty as a distinct driver of assets correlation dynamics. By embedding trade uncertainty directly into the DCC framework, the proposed specification provides a transparent empirical setting to assess how trade-policy shocks propagate to stock–T bill comovements. Moreover, the proposed specification accommodates regime-dependent transmission channels through interaction terms, enabling formal tests of whether the sensitivity of correlations to uncertainty varies across political environments.

As an empirical application, we focus on trade policy uncertainty \citep[TPU,][]{Caldara:Iacoviello:Molligo:Prestipino:Raffo:2020} as a prominent source of policy-related risk, and consider three specifications. First, we include TPU as an exogenous regressor in the correlation dynamics. Second, we introduce a political regime dummy to capture potential macroeconomic differences (e.g., inflation, monetary policy stance, financial market volatility) across administrations, even including trade-related policies. Third, we allow for interaction effects between uncertainty and the regime indicator, enabling us to assess whether the transmission of trade-related uncertainty to correlation dynamics varies across political environments, in line with evidence of regime-dependent patterns in financial market correlations \citep{Demirer:Gupta:2018}.

The empirical analysis is conducted using daily data on major U.S. stock indices—the S\&P 500, Dow Jones Industrial Average, Nasdaq Composite, and Russell 2000—and the 3-month Treasury bill (T-bill), which serves as a short-term benchmark and proxy for safe-haven assets. The results indicate that trade policy uncertainty is positively associated with stock-T bill correlations and that this effect is amplified in specific regimes. Moreover, uncertainty shocks are shown to affect not only the magnitude but also the sign of correlations, pointing to structural changes in the correlation-generating process, as confirmed by a structural break analysis of the estimated correlations. These findings - which are robust relative to the bond maturity as well as to the effect of financial and general policy uncertainty - highlight the importance of explicitly accounting for observable sources of uncertainty when modeling time-varying correlations and evaluating diversification properties in financial markets. In other words, TPU represents a relevant source of uncertainty for stock-T bill correlations, containing incremental information beyond general policy uncertainty. Finally, the out-of-sample forecast analysis reveals the importance of both trade uncertainty and political regime in enhancing the economic relevance of covariance forecast.

The paper is structured as follows. Section \ref{sec:model} introduces the model; Section \ref{sec:empirical_application} presents and discusses the estimation results, while Section \ref{sec:structural_break} is devoted to the evaluation of structural breaks in the Data Generating Process. Section \ref{sec:robustness} shows some robustness check analyses. Finally, Section \ref{sec:conclusion} concludes the paper with some final remarks.

\section{The Model}
\label{sec:model}

In the multivariate framework, the Dynamic Conditional Correlation (DCC) model by \cite{Engle:2002b} 
has proven effective in capturing the time-varying features of asset correlations, given its capability to address the two main problems that arise when estimating covariances or correlations. According to \citet{Bauwens:Laurent:Rombouts:2006}, these are the curse of dimensionality, which refers to the exponential growth of the number of parameters as the number of series increases and the need to guarantee positive-definiteness of the estimated covariance (correlation) matrix at the same time. \cite{Engle:2002b} proposes a simple two step procedure: in the first step conditional variances are estimated, while correlations are modeled in the second step, based on the first-step estimation results. In particular, let 
$$\mathbf{H_t} = \mathbf{S_t R_t S_t},$$
be the conditional covariance matrix of the ($N \times 1$) vector of a daily series $\bm{r}_t$, where $\mathbf{S}_t$ is a diagonal matrix of conditional standard deviations, and $\mathbf{R}_t$ is the correlation matrix.
In other words, we assume that $\bm{r}_t| F_{t-1} \sim N(0, \mathbf{H_t}), \quad t = 1, \ldots, T$, where $F_{t-1}$ denotes the available information set. 

In the first step, for each asset \textit{i}, we consider the GJR-GARCH model \citep{Glosten:Jaganathan:Runkle:1993} to obtain the conditional variance $h^2_{i,t}$, i.e., each element on the main diagonal of the $\mathbf{H_t}$. The GJR-GARCH model is expressed as in Eq. (\ref{eq:gjr_garch})

\begin{equation}
	h^2_{i,t}=\omega_i+\alpha_i r^2_{i,t-1}+\beta_i h^2_{i,t-1} +\gamma_i r^2_{i,t-1} I_{i,t-1}, \qquad i=1,\dots,N \label{eq:gjr_garch}
\end{equation}
where $I_{i,t}$ is the indicator function for negative returns: it takes the value 1 when the $i-$th element $r_{i,t}$, is negative and 0 otherwise. As usual in the GARCH-type model, $r^2_{i,t-1}$ represents the ARCH term, capturing the most recent shocks, while $h^2_{i,t-1}$ is the GARCH term that measures the impact of lagged conditional volatility. For stationarity, the condition $(\alpha_i+\beta_i+\frac{\gamma_i}{2})<1$ is imposed, while the constraints $\omega_i>0,\alpha_i\ge 0,\beta_i\ge 0,\gamma_i\ge 0$ ensure positivity of $h^2_{i,t}$.

Under the assumption of normality for the innovation term, the log-likelihood for the GJR--GARCH model in Eq. (\ref{eq:gjr_garch}) is given by
\begin{equation}
	\ell(\vartheta) = -\frac{1}{2}\sum_{t=1}^{T}
	\left[
	\log(2\pi)
	+ \log h_{i,t}^2
	+ \frac{r_{i,t}^2}{h_{i,t}^2}
	\right],
	\label{eq:loglik_gjr}
\end{equation}
where $\vartheta = (\omega_i,\alpha_i,\beta_i,\gamma_i)'$ denotes the parameter
vector and $h_{i,t}^2$ is computed recursively according to Eq. (\ref{eq:gjr_garch}).

In the second step, the procedure estimates $\mathbf{R}_t$ using the so-called de-GARCHed returns,
\begin{equation}
	\tilde{\epsilon}_{i,t} = r_{i,t}/h_{i,t}
	\label{eq:stand_res}
\end{equation}
which are obtained from the first-step variance estimation. More in detail, in the specification proposed by \citet{Engle:2002b} the correlation process has a GARCH-type structure, where the conditional correlations depend on their own lags and on lagged products of de-GARCHed returns, representing the impact of recent market shocks. The DCC model is specified as in Eq. (\ref{eq:dcc_full})

\begin{align}
	\mathbf{R}_t &= \tilde{\mathbf{Q}}_t^{-1}\mathbf{Q}_t\tilde{\mathbf{Q}}_t^{-1}, 
	\quad \tilde{\mathbf{Q}}_t = \mathrm{diag}(\sqrt{Q_{11,t}}, \ldots, \sqrt{Q_{nn,t}}) \notag\\
	\mathbf{Q}_t &= (1-\theta_1-\theta_2)\bar{\mathbf{R}} + \theta_1\tilde{\mathbf{Q}}_{t-1}\tilde{\epsilon}_{t-1}\tilde{\epsilon}_{t-1}^\prime\tilde{\mathbf{Q}}_{t-1} + \theta_2\mathbf{Q}_{t-1} \label{eq:dcc_full}
\end{align}
where $\bar{\mathbf{R}}$ is the unconditional (sample) correlation matrix. The second term, with coefficient $\theta_1$, depends on the lagged outer product of the de-GARCHed returns, capturing the impact of market news; finally, coefficient $\theta_2$ captures the autoregressive component.

In its standard form, the DCC accounts for the impact of exogenous shocks only indirectly. The term
$\tilde{\mathbf{Q}}_{t-1}\tilde{\epsilon}_{t-1}\tilde{\epsilon}_{t-1}^\prime\tilde{\mathbf{Q}}_{t-1}$ 
captures the effect of past market news and, by doing so, implicitly summarizes the influence of all shocks occurring on a given trading day. In this paper, we employ the DCC model with exogenous variables (DCC-X) to directly model the impact of trade policy uncertainty on the correlation between stock indices and the 3-month Treasury bill (T-bill). Choosing the T-bill allows us to focus on very short-term market reactions and to evaluate its role as a risk-free asset, particularly in the context of uncertain or inconsistent trade policy decisions.

Zooming in on the conditional correlation equation, the DCC-X model is defined as in Eq. (\ref{eq:dccx}):
\begin{equation}
	\bm{Q}_t = (1 - \theta_1 - \theta_2-\theta_3\bar{x})\bar{\bm{Q}} + \theta_1\tilde{\mathbf{Q}}_{t-1}\tilde{\epsilon}_{t-1}\tilde{\epsilon}_{t-1}^\prime\tilde{\mathbf{Q}}_{t-1} + \theta_2\mathbf{Q}_{t-1}+\theta_3 x_{t-1}, \label{eq:dccx}
\end{equation}
where $x_{t-1}$ is an exogenous variable accounting for trade policy uncertainty. Our model, the GJR-DCC-X, resembles the Volatility Dependent Conditional Correlation \citep[VDCC,][]{Bauwens:Otranto:2016}, where volatility proxies are inserted as regressors in the correlations dynamics.  In the empirical application, we consider three specifications: first, including the Trade Policy Uncertainty index \citep[TPU,][]{Caldara:Iacoviello:Molligo:Prestipino:Raffo:2020},  which measures the perceived uncertainty related to U.S. trade policy; second, incorporating a dummy variable equal to 1 during the Republican administration and 0 otherwise; and third, adding an interaction term between TPU and the administration dummy. Given that the considered exogenous variables are available at a daily frequency, the DCC-X can be adopted in its simplest form, avoiding the need for more complex models such as the DCC-MIDAS \citep{Colacito:Engle:Ghysels:2011}, which are designed to handle mixed-frequency regressors.

In Eq. (\ref{eq:dccx}), $\theta_1$ and $\theta_2$
are specified as scalars rather than matrices to reduce dimensionality and ensure positive definiteness of $\bm{Q}_t$. This contrasts with both the unrestricted DCC and the BEKK \citep{Engle:Kroner:1995} models, where the use of full matrices leads to a large number of parameters, making estimation more computationally intensive and less feasible in high-dimensional settings.\footnote{An alternative approach is found in \cite{Bauwens:Otranto:2023} who employ the Hadamard-exponential operator to address both dimensionality and positive-definiteness constraints.} Furthermore, while stationarity and positiveness conditions are stated in \cite{Engle:2002b}, we expect $\theta_3$ to be strictly positive, which -- given the positiveness of $x_{t-1}$ -- is consistent with $\bm{Q}_t$ being positive-definite. In particular, if $\theta_3>0$, correlations increase with $x_{t-1}$. This is consistent with the view that rising trade-related uncertainty heightens systematic risk and jointly affects both treasury and equity markets, leading to stronger cross-asset co-movement. It is worth noting that if $\theta_3=0$ the model reduces to the standard DCC, while with $\theta_1=\theta_2=\theta_3=0$ the simplest Constant Conditional Correlation \citep[CCC,][]{Bollerslev:1990} model is obtained.

In all the considered specifications, by assuming conditional normality, correlations are estimated by Maximum Likelihood, with log-likelihood expressed as
\begin{equation}
	L(\bm{\theta}) = -\frac{TN}{2}log(2\pi) -\frac{T}{2} \log\left( |\mathbf{R_t}| \right) 
	-\frac{1}{2} \sum_{t=1}^T 
	\tilde{\bm{\epsilon}}_{t,\cdot} \, 
	\mathbf{R_t}^{-1} \, 
	\tilde{\bm{\epsilon}}_{t,\cdot}^\prime, \label{eq:loglik}
\end{equation}
where $\bm{\theta}$ denotes the parameter vector and $\tilde{\bm{\epsilon}}_{t,\cdot}$ the $t$-th row of the matrix of de-GARCHed returns. Finally we rely on robust standard errors \citep{White:1980} to shield against potential heteroskedasticity.\footnote{As an alternative, given the empirical evidence of excess kurtosis in financial returns, the model can also be estimated under a Student-$t$ distribution. The results (available upon request) show that this alternative specification does not materially affect the estimated coefficients.}

\section{Empirical Application}
\label{sec:empirical_application}
We aim to investigate the impact of trade uncertainty on correlations between stocks and treasury markets. The empirical analysis is based on daily series for the S\&P 500, Dow Jones Industrial Average, Nasdaq Composite, Russell 2000 (as representatives of the stock markets), and the US 3-month Treasury bill (T-bill) for the short term government debt market. The data set covers the period from January 2, 2015 to April 30, 2025 and is provided by Yahoo Finance. For the stock markets, we construct daily log returns from the observed price series. For the Treasury market, we use the annualized 3‑month US Treasury bill yield and compute its first differences to obtain a stationary series of daily yield changes expressed in basis points. As a measure of trade uncertainty we rely on the Trade Policy Uncertainty index \citep{Caldara:Iacoviello:Molligo:Prestipino:Raffo:2020}.\footnote{Data are available at \url{https://www.matteoiacoviello.com/tpu.htm}.}

Table 1 reports the descriptive statistics for the considered series. As expected, stock indices returns have higher means and standard deviations compared to T-bills. The skewness is negative for all stock indices, with negative returns (crashes) that occur more frequently than equivalent positive returns. Conversely, T-bill returns show a positive skewness, likely due to occasional upward adjustments in the market or interest rate curve. Finally, all series display signs of fat tails, as highlighted by the high kurtosis values, confirming the presence of extreme events beyond what would be expected under a normal distribution.

\begin{table}[t]
	\caption{Descriptive statistics. Sample period: January 2, 2015 -- April 30, 2025.}\label{tab:descriptive statistics}
	\begin{adjustbox}{max width=0.9\linewidth,center}
		\begin{tabular}{@{}lcccccc@{}}
			\toprule
			& $S\&P~500$     & $Dow~Jones$   & $Nasdaq~Composite$ & $Russell~2000$            & $T-bill$      & TPU     \\
			\midrule
			Mean               & 0.0384  & 0.0318  & 0.0503  & 0.0190  & 0.0016 & 0.1105   \\
			Standard Deviation & 1.1526  & 1.1242  & 1.3856  & 1.4673  & 0.0292  & 0.1541 \\
			Skewness           & -0.6510 & -0.8337 & -0.4485 & -0.8029 & 0.8510 & 5.2799 \\
			Kurtosis           & 18.6055 & 24.7153 & 11.8188 & 13.8810 & 21.7477 & 43.6080\\[1mm]
			\bottomrule
		\end{tabular}
	\end{adjustbox}
\end{table}
\begin{figure}[h!]
	\centering
	{\includegraphics[height=6cm,width=13cm]{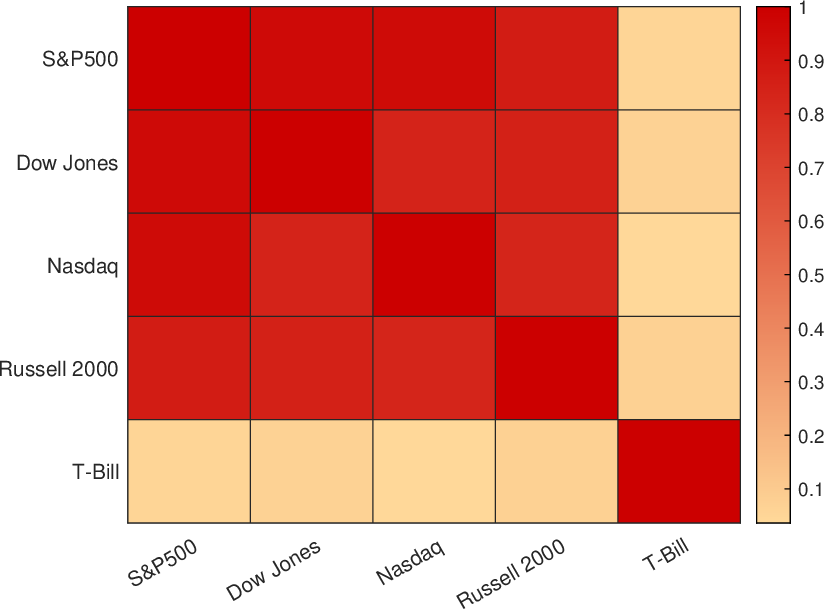}}
	
	\caption{Heatmap representing the sample correlations of the de-GARCHed returns as defined in Eq.~\ref{eq:stand_res}. Darker colors indicate stronger positive correlations, while lighter colors indicate weaker or negative correlations. }
	\label{fig:correlation_mat}
\end{figure}

Figure \ref{fig:correlation_mat} shows the sample correlation matrix of the de-GARCHed returns in Eq. \ref{eq:stand_res}. As expected, correlations among stock indices are close to 1, while correlations between stock markets and T Bills are lower, confirming the latter's role as effective instruments for mitigating portfolio risk. However, despite the de-GARCHing process, the heatmap reveals that correlations among asset classes persist, indicating that some cross-asset dependencies remain even after removing conditional volatility. This suggests the presence of common factors or residual systemic risk not captured by univariate volatility models.

\begin{figure}[h!]
	\centering
	\subfigure[Tbill vs S\&P 500]{\includegraphics[height=5cm,width=14cm]{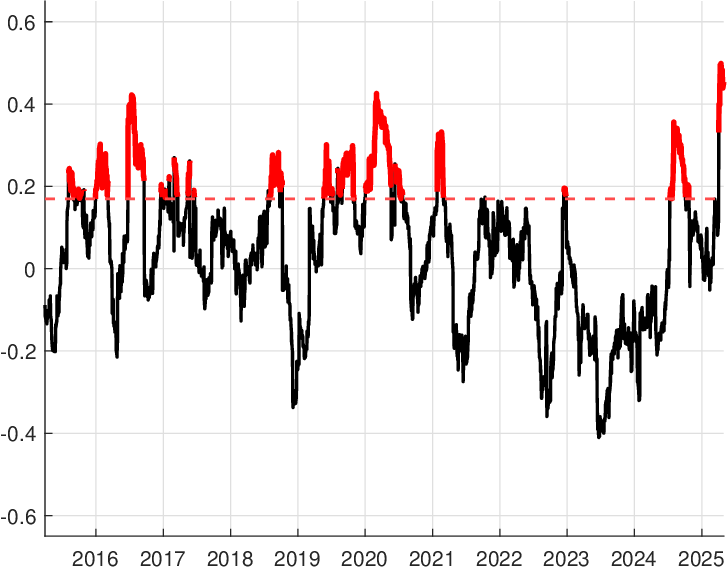}}
	
	\subfigure[Tbill vs Nasdaq]{\includegraphics[height=5cm,width=14cm]{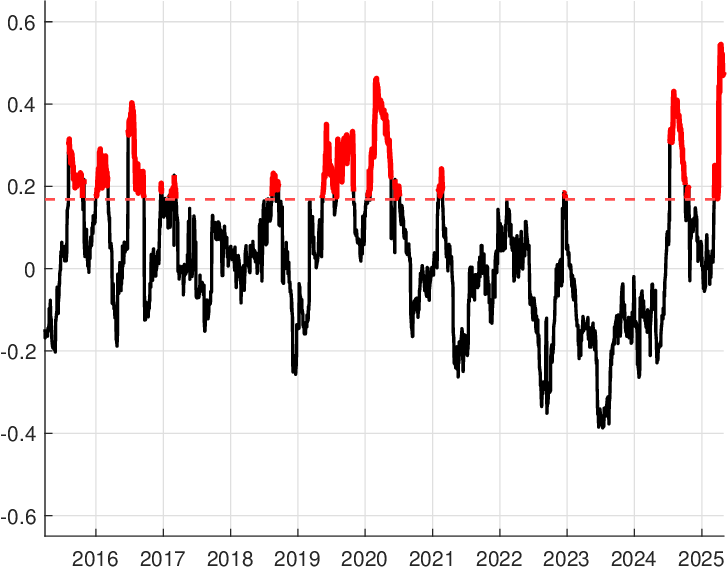}}
	
	\caption{60-day rolling correlation between stock and T bill returns (black line), with periods exceeding one standard deviation highlighted in red. The dashed red line represents the threshold of one standard deviation.}
	\label{fig:correlation_moving}
\end{figure}
This is also illustrated in Figure \ref{fig:correlation_moving}, which depicts the 60-day stock-T bill rolling correlation relative to S\&P500 (black line, panel a) and Nasdaq (panel b), with periods where correlations exceed the threshold of one standard deviation (dashed red line) highlighted in red. The time-varying nature of the correlation is apparent: the average correlation is close to zero, but there are periods of positive and high correlation, reaching up to 0.6. These episodes coincide with intervals in which the correlation exceeds the threshold of one standard deviation and are highlighted in red in the figure. Notably, these high-correlation periods align with phases of elevated uncertainty surrounding trade policies and stronger protectionist measures.

\begin{figure}[t]
	\centering
	\subfigure[Tbill vs S\&P 500]{\includegraphics[height=5cm,width=12cm]{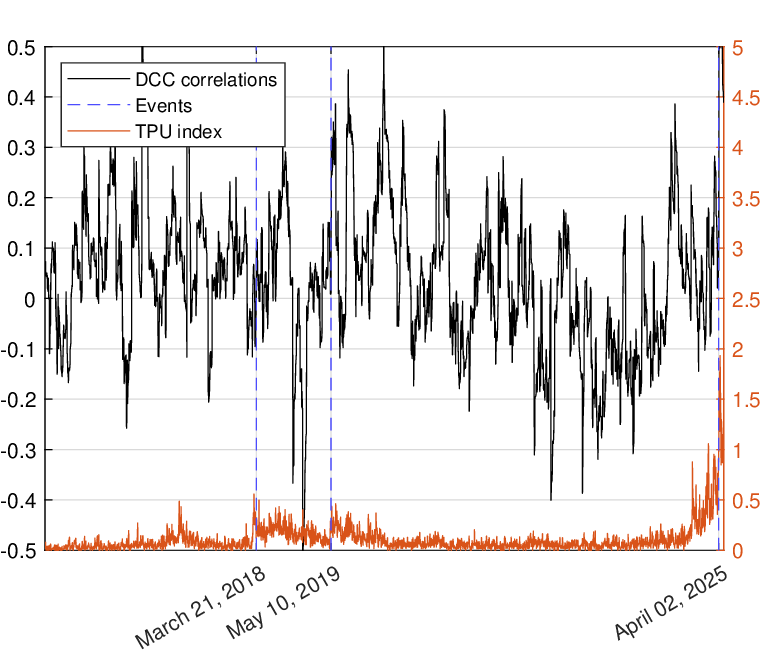}}
	
	\subfigure[Tbill vs Nasdaq]{\includegraphics[height=5cm,width=12cm]{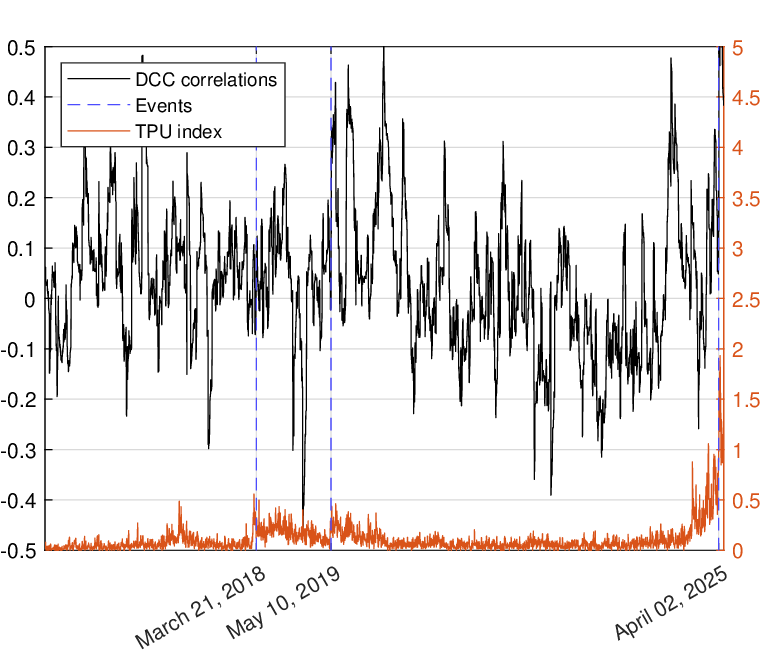}}
	
	\caption{Conditional correlations from the DCC model (black line, left axis), Trade Policy Uncertainty (TPU) index (red line, right axis), and key trade-related events (vertical blue-dashed lines)}
	\label{fig:correlation_tpu}
\end{figure}

Finally, Figure \ref{fig:correlation_tpu} displays the evolution of the conditional correlations (black line, left axis) for T bill-S\&P500 (panel a) and T bill–Nasdaq (panel b), estimated using the standard DCC model, and the TPU (red line, right axis). We observe a clear pattern of shifting between periods of positive and negative correlations, with turning points (from negative to positive values) corresponding to periods when the TPU increases. Figure \ref{fig:correlation_tpu} also shows some important dates (vertical blue dashed line) related to tariffs announcements: March 21, 2018, when the US administration decided on 25\% tariffs on imported steel and 10\% on imported aluminum; similarly, on May 10, 2019 there was an increase in tariffs against China; finally, on April 02, 2025 ``a universal baseline tariff of 10\% on all imports into the United States'' was announced. In all cases, we observe an increase in the TPU index -- which also becomes more volatile -- as well as an increase in correlations which remain positive in the immediate periods following the shock, before turning negative once the shock is absorbed. This pattern is consistent with the idea that episodes of trade-related uncertainty trigger a generalized deterioration in risk sentiment, prompting investors to reassess the riskiness of both equities and very short-term government securities. As a result, the traditional safe haven role of T Bills weakens, leading to a temporary co-movement between the two markets. Interestingly, the TPU series shows signs of clustering, with periods of low and less volatile uncertainty at the beginning of the sample and during the 2021--2025 period. Conversely, higher levels of TPU are observed during the last Republican administration. This is consistent with the broader literature linking policy-related uncertainty to financial market volatility and co-movements, such as \citet{Pastor:Veronesi:2012}, who show that increases in political uncertainty can significantly affect asset valuations and risk premia. {In this sense, the visual evidence suggests that, during the Republican administration's periods, the prevailing macroeconomic conditions are associated to increased trade policy uncertainty and higher stock-T bill co-movements. Overall, the figures provide a strong justification for augmenting the standard DCC model with exogenous variables capturing policy-driven shocks.

Table 2 reports the estimation results for the GJR-DCC-X model. Regarding the variance equation (panel a), for the stock indices, the coefficient $\hat{\alpha}$ is significant only for the Dow Jones (at a 10\% significance level) and Russell 2000 (at 5\%); $\hat{\alpha}$ is higher for the treasury market, indicating the crucial role of news in generating volatility. Furthermore, we find evidence supporting the asymmetric effect of negative returns, with $\hat{\gamma}$ being significant in all the cases. Finally, $\hat{\beta}$ is greater than 0.8, meaning that volatility is primarily driven by its past values. These features reinforce the idea that periods of negative market performance amplify volatility, which may then transmit to correlations through the DCC structure.

\begin{table}[h!]
	\caption{Estimated coefficients for the GJR model (panel a) and the DCC-X specifications (panel b). Robust standard errors (White, 1980) are reported in parentheses. Ljung–Box p-values (Ljung and Box, 1978) for first-order residual autocorrelation are reported in panel c. Sample period: January 2, 2015 -- April 30, 2025.}\label{tab:results}
	\begin{adjustbox}{max width=1\linewidth,center}
		\begin{tabular}{@{}lccccc@{}}
			\toprule
			panel a)               & $S\&P~500$     & $Dow~Jones$   & $Nasdaq$ & $Russell~2000$            & $T-bill$           \\
			\midrule
			$\omega$   & 0.0408    & 0.0400        & 0.0473             & 0.0451                     & 0.0000          \\
			& (0.0097)  & (0.0088)      & (0.0162)           & (0.0162)                   & (0.0000)        \\
			$\alpha$   & 0.0502    & 0.0484        & 0.0206             & 0.0410                     & 0.1617          \\
			& (0.0327)  & (0.0261)      & (0.0195)           & (0.0145)                   & (0.0073)        \\
			$\beta$    & 0.8029    & 0.8032        & 0.8671             & 0.8794                     & 0.8520          \\
			& (0.0309)  & (0.0300)      & (0.0315)           & (0.0244)                   & (0.0041)        \\
			$\gamma$   & 0.2412    & 0.2326        & 0.1699             & 0.1145                     & 0.0535          \\
			& (0.0503)  & (0.0422)      & (0.0478)           & (0.0273)                   & (0.0100)        \\[1mm]
			\midrule
			panel b) & $DCC$         & $DCC\text{-}X_{TPU}$ & $DCC\text{-}X_{Dummy}$    & $DCC\text{-}X_{TPU\times Dummy}$& $DCC\text{-}X_{Full}$\\
			\midrule
			$\theta_1$ & 0.0486       & 0.0464       & 0.0479       & 0.0466       & 0.0454       \\
			& (0.0051)     & (0.0049)     & (0.0050)     & (0.0050)     & (0.0051)     \\
			$\theta_2$ & 0.9294       & 0.9292       & 0.9282       & 0.9289       & 0.9297       \\
			& (0.0092)     & (0.0094)     & (0.0093)     & (0.0095)     & (0.0095)     \\
			$\theta_3$ &              & 0.0250       & 0.0040       & 0.0347       & 0.0042      \\
			&              & (0.0092)     & (0.0019)     & (0.0113)     & (0.0145)     \\
			$\theta_4$ &              &              &              &              & -0.0026      \\
			&              &              &              &              & (0.0027)     \\
			$\theta_5$ &              &              &              &              & 0.0421       \\
			&              &              &              &              & (0.0209)     \\[2mm]
			AIC        & -2511.02 & -2530.04  & -2519.97 & -2534.70 & -2532.54 \\
			BIC        & -2499.30 & -2512.46 & -2502.39 & -2517.11 & -2503.23 \\ 	LR&	&21.02 &	10.95 &	25.68 &	27.52 \\
			&	&0.00 &	0.00 &	0.00 &	0.00\\[1mm]
			\midrule
			panel c) & \multicolumn{5}{c}{Ljung–Box p-values for first-order residual autocorrelation}\\[2mm]
			$S\&P~500$   & 0.426 & 0.426 & 0.426 & 0.426 & 0.426 \\
			$Dow~Jones$   & 0.001 & 0.001 & 0.001 & 0.001 & 0.001 \\
			$Nasdaq$  & 0.390 & 0.403 & 0.413 & 0.418 & 0.412 \\
			$Russell~2000$   & 0.894 & 0.904 & 0.874 & 0.890 & 0.905 \\
			$T--Bill$ & 0.028 & 0.027 & 0.028 & 0.027 & 0.027\\
			\bottomrule
		\end{tabular}
	\end{adjustbox}
\end{table}

As shown in Figure \ref{fig:correlation_tpu}, periods of heightened trade uncertainty coincide with the implementation of new tariffs and are associated with higher correlations. In this respect,  panel b) of Table 2 shows estimation results for the DCC equations (Eq. (\ref{eq:dcc_full}) and (\ref{eq:dccx})). As for the standard DCC (column 1), as expected, $\hat\theta_1$ is approximately 0.05, indicating a moderate impact of news on correlations; by contrast, $\hat\theta_2$ is 0.929, reflecting the higher persistence feature of correlations. As regards the DCC-X, when considered in isolation, the three exogenous variables enter the model with a positive sign and are highly significant. In detail, on average, a marginal increase in TPU is associated with a 0.025 increase in the stock-T bill correlation (column $DCC-X_{TPU}$); the correlations also increase during the Republican administration (column $DCC-X_{Dummy}$), with the dummy coefficient equal to 0.004 and significant at a 5\% level. This aligns with the findings of \cite{Demirer:Gupta:2018}, who provide evidence of a presidential cycle effect on the correlation of financial returns, which decreases and becomes negative during Democratic administrations. These results are also consistent with the broader evidence that economic and policy uncertainty affects the joint dynamics of financial markets. For instance, \citet{Bekaert:Hoerova:LoDuca2013} show that uncertainty shocks (in a broader sense) can raise risk premia and alter the covariance structure across assets. In our context, trade policy uncertainty appears to operate as a similar mechanism: by increasing macro-financial uncertainty, it modifies the conditional correlation between stocks and short-term government securities

Figure \ref{fig:irf} shows the Impulse Response Function, IRF (black line, left axis), expressed in percentage points, of the stock-T bill correlation following a one-standard-deviation shock in TPU. This allows us to measure the additional effect of the shock relative to the baseline scenario, where the baseline corresponds to the correlation dynamics in the absence of any shock (i.e., starting from the average level of correlations). One day after the shock, the correlation increases by 0.4 percentage points, compared to the baseline scenario. The IRF gradually decays toward zero, taking approximately 60 days to be fully absorbed, which reflects the high persistence of correlations estimated in the DCC-X model. To benchmark the magnitude of the TPU shock, we compare this effect with that generated by a financial volatility shock proxied by the VIX index (red line, right axis). A one-standard-deviation increase in the VIX produces a smaller response (in absolute terms) of about 0.06 percentage points. However, this difference partly reflects the higher volatility of TPU relative to the VIX series in our sample: when normalizing the responses by the standard deviation of the underlying variables, the implied marginal sensitivity of correlations to financial market volatility remains stronger per unit change (2.60 and 8.79 for TPU and VIX, respectively). This suggests that TPU shocks generate economically meaningful responses but this impact is lower than that associated to a volatility shock.

\begin{figure}[t]
	\centering
	\subfigure[Tbill vs S\&P 500]{\includegraphics[height=4cm,width=6.7cm]{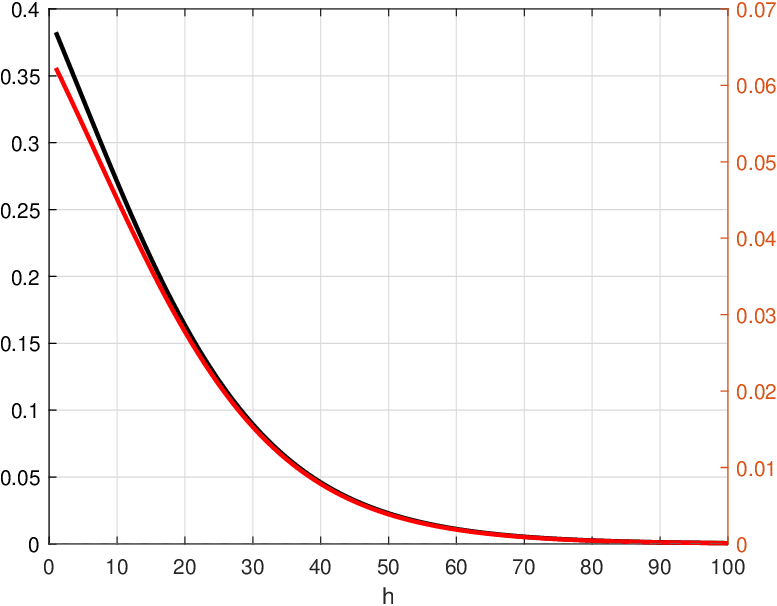}}
	\subfigure[Tbill vs Dow Jones]{\includegraphics[height=4cm,width=6.7cm]{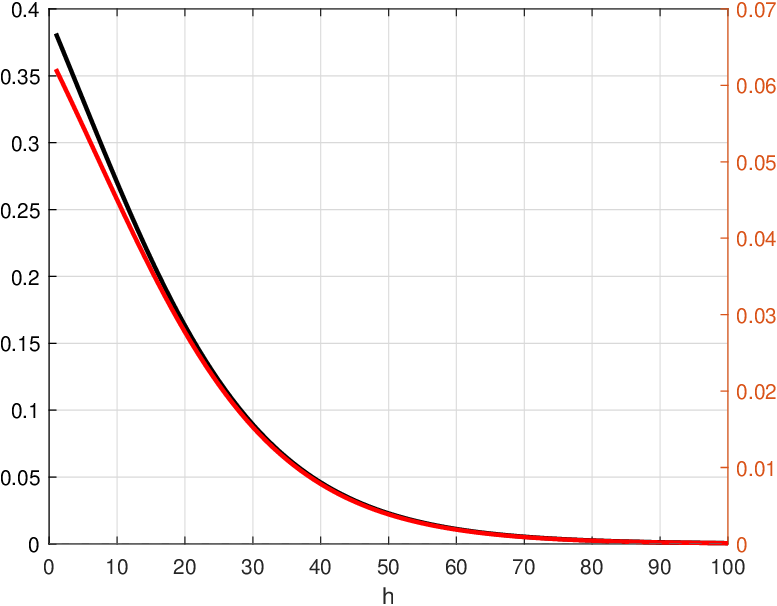}}\\
	
	\subfigure[Tbill vs Nasdaq]{\includegraphics[height=4cm,width=6.7cm]{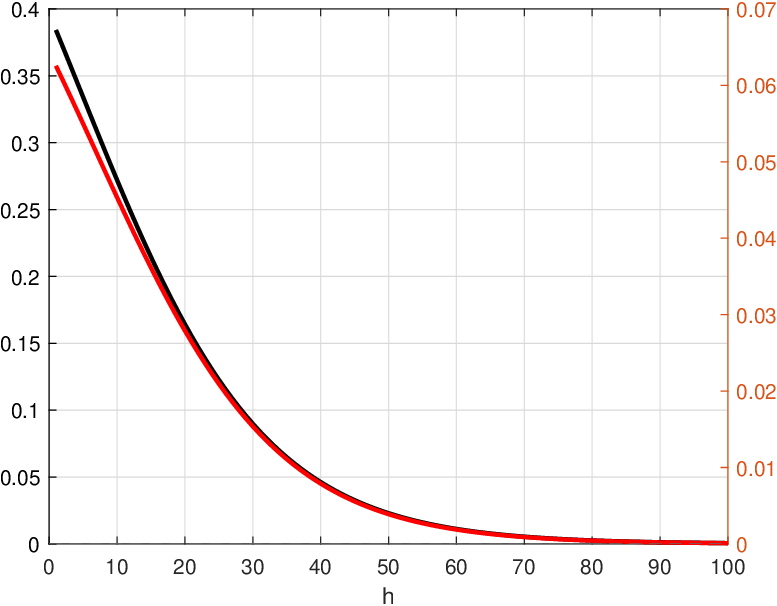}}
	\subfigure[Tbill vs Russell 2000]{\includegraphics[height=4cm,width=6.7cm]{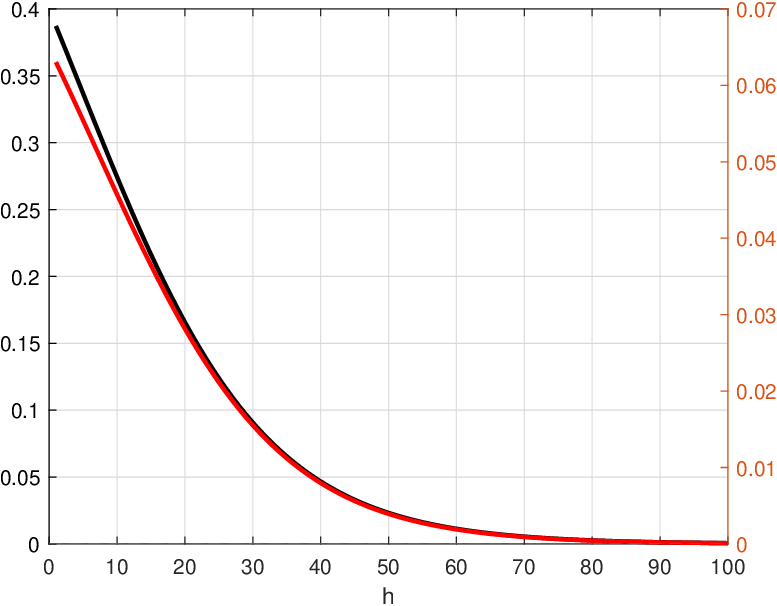}}
	
	\caption{Impulse Response Function (IRF) of the stock-T bill correlation to a one-standard-deviation shock in trade policy uncertainty (black line, left axis) and VIX (red line, right axis), expressed in percentage points}
	\label{fig:irf}
\end{figure}

Finally, the positive and statistically significant coefficient related to the interaction term reveals that trade policy uncertainty exerts a larger influence on stock-T bill correlations during periods in which the US administration pursued a more interventionist and unpredictable trade agenda. This suggests the presence of a regime-dependent transmission mechanism whereby the economic and financial consequences of trade-related news are amplified when policy actions become more frequent, abrupt, or difficult to forecast.  Recently, the US administration adopted a sequence of tariff announcements, countermeasures, and negotiation stand-offs, contributing to heightened global trade tensions. Such an environment likely heightened investors' sensitivity to trade uncertainty, leading to a stronger alignment of movements across asset classes. From a market perspective, this implies that uncertainty shocks were interpreted not as transitory deviations but as signals of a potentially persistent shift in trade policy.

For completeness, we have also considered a full specification including all the variables as regressors (column $DCC\text{-}X_{Full}$). The parameters $\theta_1$ and $\theta_2$ remain largely unchanged, while the only significant coefficient is associated with the interaction term, suggesting that the influence of uncertainty becomes more pronounced under the Republican administration. However, since this specification likely suffers from multicollinearity,\footnote{As a matter of fact, the correlation between TPU and the interaction term is 0.913.} we prefer to include each exogenous variable separately. This mitigates multicollinearity concerns and allows us to isolate the individual impact of each variable on the estimated correlation. This modeling choice is further supported by the Akaike \citep[AIC,][]{Akaike:1974} and Bayesian \citep[BIC,][]{Schwarz:1978} information criteria reported in the last two rows of Table 2, which indicate that the preferred model is $DCC\text{-}X_{TPU \times Dummy}$. Additionally, while the standard DCC model exhibits the lowest fit quality according to AIC and BIC, the Full model is preferred only relative to $DCC\text{-}X_{Dummy}$. Likelihood Ratio (LR) tests further support the superiority of DCC-X models, rejecting the null hypothesis that the standard DCC (the restricted model) provides an adequate fit for the data in all specifications.

Residual diagnostics based on standardized residuals (Table 2, Panel c) indicate that the model adequately captures serial dependence at lag 1. In particular, the null hypothesis of no residual autocorrelation in the Ljung–Box test \citep{Ljung:Box:1978} cannot be rejected at the 1\% significance level for all the series, with the only exception being the Dow Jones index.

The economic implications of these findings are substantial. Episodes of heightened trade uncertainty reduce the diversification benefits traditionally offered by T Bills, as correlations become positive precisely when investors typically seek safe assets. This correlation breakdown reduces the hedging value of short-term government securities and may contribute to liquidity-driven sell-offs (i.e., liquidation cascades) as investors rebalance portfolios toward liquidity rather than safety. Such dynamics highlight the importance of incorporating policy-related uncertainty into correlation models, particularly when evaluating short-term risk management strategies.

\subsection{Out-of-sample Forecast evaluation}
\label{sec:oos}
The forecasting capability of the considered specifications is evaluated by estimating the models over an in-sample period spanning from January 2, 2015 to December 30, 2022, and generating forecasts of conditional variances and correlations for the out-of-sample period from January 3, 2023 to April 30, 2025, resulting in a total of 583 forecasts.

Model comparison is conducted using the Model Confidence Set \citep[MCS,][]{Hansen:Lunde:Nason:2011} procedure, which identifies the set of models with superior predictive ability for a given significance level and loss function. We rely on several loss functions. As statistical loss functions we consider the Frobenius norm loss (F) function
	\begin{equation*}
		F_{\tau}=\sum_{i=1}^{N}\sum_{j=1}^{N}
		\left(\hat{\sigma}_{ij,\tau}-r_{i,\tau}r_{j,\tau}\right)^2
	\end{equation*}
	and the QLike loss
	\begin{equation*}
		QLike_{\tau}=
		\left(
		\log|\hat{\mathbf{H}}_{\tau}|
		+
		\mathrm{tr}\left(
		\hat{\mathbf{H}}_{\tau}^{-1}\mathbf{C}_{\tau}
		\right)
		\right),
	\end{equation*}
	where $\hat{\mathbf{H}}_{\tau}$ denotes the forecasted covariance matrix, $\mathbf{C}_{\tau}=\mathbf{r}_{\tau}\mathbf{r}_{\tau}'$ is the realized covariance proxy, and $T_h$ denotes the number of forecasts.

To evaluate whether differences in correlation forecasts translate into economically meaningful improvements in portfolio risk, we also consider economic loss functions, such as the Global Minimum Variance ($GMV$) portfolio \citep{Engle:Colacito:2006}
	\begin{equation}
		GMV_{\tau}=
		\left(
		\hat{\mathbf{v}}_{\tau}'\hat{\mathbf{H}}_{\tau}\hat{\mathbf{v}}_{\tau}
		\right),
		\label{eq:gmv}
	\end{equation}
	with
	\begin{equation}
		\hat{\mathbf{v}}_{\tau}=
		\sqrt{n}
		\frac{\hat{\mathbf{H}}_{\tau}^{-1}\mathbf{j}_n}
		{\mathbf{j}_n'\hat{\mathbf{H}}_{\tau}^{-1}\mathbf{j}_n},
	\end{equation}
	where $\mathbf{j}_n$ denotes an $n\times1$ vector of ones. We also consider the Realized Portfolio Variance (RPV)
	\begin{equation}
		RPV_{\tau}=
		\left(
		r_{p,\tau}-\bar r_p
		\right)^2,
	\end{equation}
	where $r_{p,\tau}=\hat{\mathbf{v}}_{\tau}'\mathbf{r}_{\tau}$ denotes the realized portfolio return.

Table 3 reports the MCS $p$-values computed using the range statistic $T_R$. Based on the Frobenius loss, all the considered specifications enter the set of superior models, with the full model emerging as the best-performing specification ($p$-value = 1). Based on the QLike loss, which is consistent according to \citet{Patton:2011}, the standard DCC model performs best, although the $DCC\text{-}X_{TPU}$ specification exhibits statistically equivalent forecasting ability. Interestingly, when economic loss functions are considered, the results change. At the 5\% significance level, the standard DCC model is always excluded from the superior set, while the $DCC\text{-}X_{Dummy}$ and $DCC\text{-}X_{TPU\times Dummy}$ specifications emerge as the best-performing models. Finally, according to the RPV loss, the full model is the only specification included in the superior set. These findings indicate that augmenting the DCC framework with trade-policy-related variables improves the economic relevance of correlation forecasts. While statistical loss functions provide mixed evidence, portfolio-based evaluation consistently favors models that incorporate political regimes and trade policy uncertainty. In particular, the exclusion of the standard DCC model when economic loss functions are considered highlights the role of regime-specific effects in shaping stock--T bill correlation dynamics. Overall, these results suggest that trade policy uncertainty and political regimes provide useful information for forecasting cross-asset comovements.

\begin{table}[h!]
	\caption{P-values for the MCS out-of-sample forecasting evaluation based on the $T_R$ statistics. Loss function:  Frobenius norm loss(F), QLike, Global Minumum Variance (GMV) and Realized Portfolio Variance (RPV). Estimation period: January 2, 2015 -- December 30, 2022. Forecasting period: January 3, 2023 -- April 30, 2025.}\label{tab:mcs}
	\begin{adjustbox}{max width=1\linewidth,center}
		\begin{tabular}{@{}lcccc@{}}
			\toprule
			& F & QLike                         & GMV                           & RPV    \\
			\midrule
			$DCC$                     & 0.9561         & 1.0000                        & 0.0438 & 0.0336 \\
			$DCC-X_{TPU}$             & 0.9561         & 0.7618                        & 0.0438 & 0.0336 \\
			$DCC-X_{Dummy}$           & 0.9780         & 0.0000 & 0.2626 &  0.0336 \\
			$DCC-X_{TPU\times Dummy}$ & 0.9561         & 0.0000 & 1.0000 & 0.0336 \\
			$DCC-X_{Full}$            & 1.0000         & 0.0000 & 0.0438 &  1.0000 \\[1mm]
			\bottomrule
		\end{tabular}
	\end{adjustbox}
\end{table}

\section{Detecting Structural Shifts in Trade Policy Impact}
\label{sec:structural_break}
In the empirical application, we have used the TPU index as a proxy for uncertainty, demonstrating that it significantly affects correlations, with the effect being amplified by protectionist measures, such as the establishment of tariffs. To further investigate the impact of specific trade policy events, we test for structural breaks in the correlation dynamics around the key tariffs announcement dates highlighted in Figure \ref{fig:correlation_tpu}, focusing in particular on potential shifts in the $\theta_3$ parameter.\footnote{Due to insufficient post-event observations, the procedure could not be applied to the third date, April 02, 2025.} This approach allows us to examine whether the sensitivity of correlations to trade policy uncertainty changes discretely after major protectionist shocks, rather than evolving smoothly over time. Specifically, we estimate the following model:

\begin{equation}
	\bm{Q}_t = (1 - \theta_1 - \theta_2-(\theta_3+\delta D_t)\bar{x})\bar{\bm{Q}} + \theta_1 {y}_{t-1}  {y}_{t-1}^\top + \theta_2 \bm{Q}_{t-1}+(\theta_3+\delta D_t) x_{t-1}, \label{eq:dccx_sb}
\end{equation}
where $D_t$  is a dummy variable taking values of 1 from the announcement day onward, allowing for the effect of uncertainty on correlations to shift after the tariffs shocks. In this framework, $\delta$ captures the magnitude and direction of the parameter shift in $\theta_3$, allowing us to formally assess whether the relationship between uncertainty and correlation dynamics changes discretely following each event.

\begin{table}[h!]
	\caption{Estimated coefficients for the DCC-X specifications (Eq. 7). Robust standard errors (White, 1980) are reported in parentheses. Sample period: January 2, 2015 -- April 30, 2025.}\label{tab:structural_break}
	\begin{adjustbox}{max width=1\linewidth,center}
		\begin{tabular}{@{}lccc@{}}
			\toprule
			& \multicolumn{3}{c}{March 21, 2018} \\
			& $DCC-X_{TPU}$ & $DCC-X_{Dummy}$ & $DCC-X_{TPU\times Dummy} $ \\
			
			\midrule
			$\theta_1$ & 0.0453        & 0.0473          & 0.0453                   \\
			& (0.0048)      & (0.0050)        & (0.0050)                   \\
			$\theta_2$ & 0.9292        & 0.9300          & 0.9305     \\
			& (0.0091)      & (0.0093)        & (0.0093)                \\
			$\theta_3$ & 0.0856        & 0.0087          & 0.1023       \\
			& (0.0215)      & (0.0038)        & (0.0312)                   \\
			$\delta$   & -0.0680       & -0.0068         & -0.0751       \\
			& (0.0206)      & (0.0042)        & (0.0315)              \\
			\midrule
			& \multicolumn{3}{c}{May 10, 2019}\\
			& $DCC-X_{TPU}$ & $DCC-X_{Dummy}$ & $DCC-X_{TPU\times Dummy} $ \\
			$\theta_1$                    & 0.0468        & 0.0479          & 0.0469                     \\
			& (0.0048)      & (0.0050)        & (0.0049)                   \\
			$\theta_2$      & 0.9281        & 0.9295          & 0.9286                     \\
			& (0.0090)      & (0.0091)        & (0.0093)                   \\
			$\theta_3$   & 0.0534        & 0.0070          & 0.0485                     \\
			& (0.0168)      & (0.0027)        & (0.0198)                   \\
			$\delta$               & -0.0403       & -0.0072         & -0.0226                    \\
			& (0.0166)      & (0.0033)        & (0.0205) \\
			\bottomrule
		\end{tabular}
	\end{adjustbox}
\end{table}

Table 4 reports the estimation results for two key tariff announcement dates that would represent crucial dates for trade policy uncertainty: March 21, 2018, and May 10, 2019. In detail, we select March 21, 2018 to capture the first US tariffs on steel and aluminum, and May 10, 2019 to mark the implementation of higher tariffs on Chinese goods following failed negotiations, and reflecting a clear escalation in the US–China trade conflict. The significantly negative $\hat{\delta}$ suggests that, while trade policy uncertainty (TPU) continues to exert a positive effect on stock-T bill correlations ($\theta_3 + \delta > 0$), this impact is substantially weaker following the tariff announcements. This result indicates a discrete adjustment in the influence of TPU on stock-T bill co-movements, indicating that policy shocks alter the sensitivity of correlations rather than merely adding short-term noise. This attenuation is economically intuitive. First, this may reflect a market adjustment mechanism, whereby the initial shock from the announcement triggers a sharp reaction in correlations, but subsequent movements in TPU may be incorporated into asset prices more efficiently as investors learn about the direction and persistence of trade policy. In this sense, once major tariff actions have been revealed, additional fluctuations in TPU may carry less marginal informational content, leading to a weaker effect on correlation dynamics over time. In this respect, Figure \ref{fig:theta3_rolling} shows the estimation of $\hat{\theta}_3$ (black line), obtained through a rolling window estimation of the $DCC-X_{TPU}$ model. Specifically, we use an estimation window of 750 observations and re-estimate the model with a step size of one observation. In particular, we find that after the announcement the coefficient recorded a 51\% increment on March 21, 2018 (from 0.006 to 0.011) and a 55\% increment on May 10, 2019 (from 0.006 to 0.010). In both cases, this effect seems to be temporary. For the latter date, $\hat{\theta_3}$ returned to the pre-announcement level in approximately 30 days, before entering in a period of monotonically increment during the COVID-19 pandemic and the onset of the Russia-Ukraine war.
	
Second, TPU exhibits lower levels and reduced volatility in the post-announcement windows (see Figure \ref{fig:correlation_tpu}), indicating that spikes in uncertainty become less frequent and less pronounced, thereby attenuating its impact on correlation dynamics. Moreover, these results align with the dynamics shown earlier in Figure \ref{fig:correlation_tpu}, where increases in TPU tend to coincide with upward spikes in correlations, but these jumps are often followed by partial reversion as the market stabilizes, consistent with short-lived but strong effects of uncertainty shocks on asset co-movements.

In conclusion, tariffs announcements induce a change in the parameters governing the data-generating process of stock-T bill correlations, leading to a reassessment of investor expectations and abrupt changes in correlation dynamics. These results underscore the need to explicitly account for policy-driven structural breaks when modeling correlation dynamics, particularly in environments characterized by recurrent policy-induced uncertainty shocks.

\begin{figure}[t]
	\centering
	{\includegraphics[height=8cm,width=14cm]{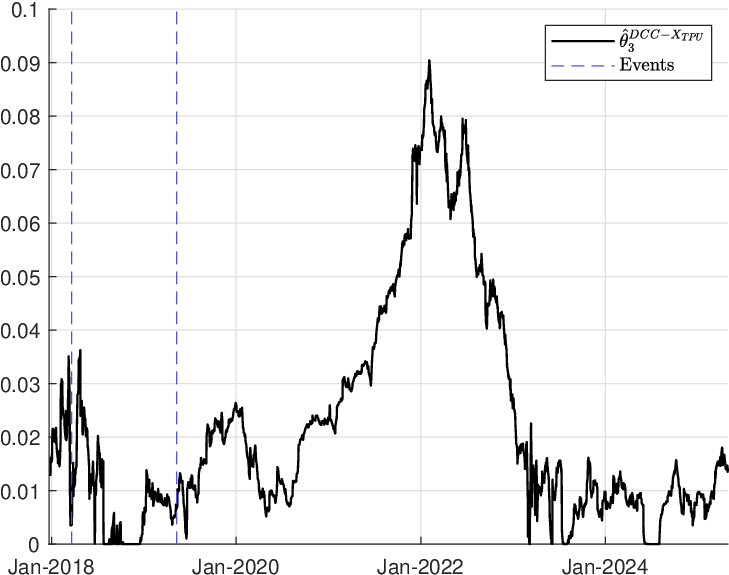}}
	
	\caption{$\hat{\theta}_3$ rolling window estimated coefficient (black line) from the $DCC-X_{TPU}$ model and break dates (dashed blue line, details in the text).}
	\label{fig:theta3_rolling}
\end{figure}

\section{Robustness Checks}
\label{sec:robustness}

In this section we present a set of robustness analyses. First, to ensure that our results are not driven by the maturity of the government security used as safe asset, we replicate the main analysis by replacing the 3-month T-bill with the 10-year Treasury bond (T-bond). This allows us to assess whether trade policy uncertainty affects stock–bond correlations also when considering medium- and long-term government securities.  Second, to demonstrate that TPU contains incremental information beyond general policy uncertainty, we augment the $DCC\text{-}X$ specification by including additional regressors capturing financial market uncertainty and general policy uncertainty. In particular, we use the Chicago Board Options Exchange Volatility Index (VIX) as a proxy for financial market uncertainty, while the Economic Policy Uncertainty index \citep[EPU,][]{Baker:Bloom:Davis:2016} is employed as a measure of general policy uncertainty.

Results for the stock-T Bond correlations are reported in Table 5, panel a). The correlation process remains highly persistent ($\hat{\theta}_2>0.9$), while recent innovations still exert a moderate effect ($\hat{\theta}_1>0.05$). Importantly, the TPU coefficient enters with a positive sign and a magnitude comparable to that obtained in the stock-T bill specification. Similar results are obtained across the alternative specifications, including the full model where the only statistically significant coefficient is associated with the interaction term.

	\begin{table}[h!]
	\caption{Panel a): Estimated coefficients for the DCC-X specifications relative to the stocks-Tbond correlations.Panel b): Estimated coefficients for the DCC-X specifications augmented with VIX and EPU. Robust standard errors (White, 1980) are reported in parentheses. Sample period: January 2, 2015 -- April 30, 2025.}\label{tab:results_bond}
	\begin{adjustbox}{max width=1\linewidth,center}
		\begin{tabular}{@{}lccccc@{}}
			\toprule
			& $DCC$         & $DCC\text{-}X_{TPU}$ & $DCC\text{-}X_{Dummy}$    & $DCC\text{-}X_{TPU\times Dummy}$& $DCC\text{-}X_{Full}$\\
			\midrule
			panel a)& \multicolumn{5}{c}{Robustness check: 10-year Treasury Bond}\\[2mm]
			$\theta_1$ & 0.0537   & 0.0513   & 0.0522    & 0.0506                          & 0.0521                          \\
			& (0.0050) & (0.0049) & (0.0049)  & (0.0050)                        & (0.0052)                        \\
			$\theta_2$ & 0.9235   & 0.9243   & 0.9214    & 0.9241                          & 0.0927                         \\
			& (0.0084) & (0.0086) & (0.0088)  & (0.0089)                        & (0.0090)                        \\
			$\theta_3$ &          & 0.0202   & 0.0066    & 0.0404                          & -0.0244                         \\
			&          & (0.0093) & (0.0024)  & (0.0124)                        & (0.0137)                        \\
			$\theta_4$ &          &          &           &                                 & 0.0020                          \\
			&          &          &           &                                 & (0.0038)                        \\
			$\theta_5$ &          &          &           &                                 & 0.0582                          \\
			&          &          &           &                                 & (0.0236)                         \\[2mm]
			\midrule
			panel b)& \multicolumn{5}{c}{Robustness check: VIX and EPU}\\[2mm]
			$\theta_1$          &     & 0.0440   & 0.0469    & 0.0441                         & 0.0439                          \\
			&     & (0.0053) & (0.0050)  & (0.0053)                       & (0.0053)                        \\
			$\theta_2$          &     & 0.9322   & 0.9283    & 0.9303                         & 0.9310                          \\
			&     & (0.0101) & (00096)   & (0.0105)                       & (0.0104)                        \\
			$\theta_3$                 &     & 0.0424   &   0.0060        &     0.0508                           & 0.0108                          \\
			&     & (0.0107) &   (0.0025)        &           (0.0120)                     & (0.0178)                        \\
			$\theta_4$                  &     &          &     &                                & 0.0000                          \\
			&     &          &  &                                & (0.0030)                        \\
			$\theta_5$    &     &          &           &                          & 0.0409                          \\
			&     &          &           &                        & (00226)                         \\
			VIX                 &     & 0.3174   & 0.3376    & 0.4946                         & 0.4466                          \\
			&     & (0.1137) & (0.1261)  & (0.1229)                       & {\color[HTML]{000000} (0.1385)} \\
			EPU                 &     & -0.0580  & -0.0497   & {\color[HTML]{000000} -0.0717} & -0.0689                         \\
			&     & (0.0107) & (0.0155)  & (0.0123)                       & (0.0126)    \\
			\bottomrule
		\end{tabular}
	\end{adjustbox}
\end{table}

Finally, Table 5, panel b, reports estimation results when the VIX and the EPU index are included in the $Q_t$ dynamics. In the $DCC\text{-}X_{TPU}$ specification (column 1), the TPU coefficient remains positive and statistically significant at the 1\% level. As expected, the VIX emerges as an important driver of stock-T bill correlations, with an estimated coefficient of 0.338. The EPU index instead enters with a negative coefficient. This result suggests that, once trade-specific uncertainty and financial market volatility are accounted for, the residual component of general policy uncertainty is associated with stronger flight-to-safety dynamics, thereby reducing stock-T bill correlations \citep[see, for example,][]{Connolly:Stivers:Sun:2005,Baele:Bekaert:Inghelbrecht:2010}. Overall, these findings indicate that TPU contains incremental information for stock-T bill correlation dynamics that is not subsumed by broader policy uncertainty or financial market volatility. Finally, the coefficient associated with the presidential dummy (column 2) remains largely unchanged after controlling for VIX and EPU. This suggests that the political regime indicator is not merely capturing variations in financial market volatility or general policy uncertainty, but reflects regime-specific effects that are orthogonal to these broader uncertainty measures.

\section{Final remarks}
\label{sec:conclusion}
This paper investigates how Trade Policy Uncertainty (TPU) influences the time-varying correlation between major U.S. stock indices and the 3-month Treasury bill (T bill), employing an augmented Dynamic Conditional Correlation (GJR-DCC-X) framework. By explicitly incorporating the TPU index and political variables into the correlation dynamics, the study contributes to the literature on policy uncertainty and financial market co-movements, addressing a gap by focusing on the asset-specific role of trade policy risk rather than on broader measures of economic policy uncertainty.

The empirical analysis yields several notable findings. Trade policy uncertainty has a statistically significant and positive effect on stock-T bill correlations, and this effect is amplified during periods characterized by a more protectionist and interventionist trade agenda, as captured by the interaction with the Presidential dummy. These results point to a regime-dependent transmission mechanism, whereby trade-related uncertainty alters cross-asset co-movements more strongly in politically unpredictable environments. As a consequence, correlations tend to increase precisely during episodes of elevated TPU, weakening the traditional safe-haven role of short-term government securities.

These findings have direct implications for portfolio allocation and risk management. When trade policy uncertainty rises, standard diversification strategies based on historically low or negative stock-T bill correlations may become less effective. Ignoring policy-driven correlation dynamics can lead to an underestimation of portfolio risk. Incorporating observable measures of policy uncertainty into correlation forecasts can therefore improve risk assessment, enhance stress-testing procedures, and support more robust portfolio construction, particularly in environments characterized by recurrent policy-induced shocks. Crucially, the impact of the TPU index is robust to the bond maturity and to the presence of both financial and general political sources of uncertainty. Furthermore, the out-of-sample forecast analysis reveals the importance of trade uncertainty in enhancing the economic relevance of correlation forecasts.

While the T bill serves as a convenient proxy for the government debt market and captures short-term dynamics, this choice limits the analysis of medium- and long-term effects of trade uncertainty, for which longer-maturity bonds would be more appropriate. Likewise, the standard DCC-X framework does not allow examination of potential regime shifts in correlations, which could be explored using more flexible models such as Smooth Transition Correlation models \citep{Silvennoinen:Terasvirta:2015}. Addressing these aspects -- by incorporating longer term bonds, higher-frequency data, or asset-specific correlation dynamics explicitly driven by trade policy uncertainty -- constitutes a promising avenue for future research.
	
	\section*{Conflict of interest disclosure} 
	\noindent No competing interest is declared.
	
	\section*{Data availability} 
	\noindent Data will be made available on request.
	
	\bibliographystyle{chicago}
	\bibliography{biblio}

\end{document}